\documentclass[%
reprint,
superscriptaddress,
 amsmath,amssymb,
 aps,
pra,
nofootinbib
]{revtex4-2}

\usepackage{physics}
\usepackage[clock]{ifsym}
\usepackage{natbib}
\usepackage{amsmath,amssymb}
\usepackage{graphicx}
\usepackage[dvipsnames]{xcolor}
\usepackage[super]{nth}
\usepackage[caption=false]{subfig}
\usepackage{float}
\usepackage[pdftex]{hyperref}
\usepackage[normalem]{ulem}
\usepackage{comment}
\newcommand{\h}{\hat }
\newcommand{\vt}{\vphantom{\frac{1}{\sqrt{2}}}}

\newcommand{\mrm}[1]{\mathrm{#1}}
\usepackage{amsmath,mathtools}

\makeatletter
\def\maketitle{
\@author@finish
\title@column\titleblock@produce
\suppressfloats[t]}
\makeatother

\begin{document}

\title{
Quantum signatures of proper time in optical ion clocks}

\author{Gabriel Sorci$^\dagger$}
\affiliation{Department of Physics, Stevens Institute of Technology, Hoboken, NJ 07030, USA}
\author{Joshua Foo$^\dagger$}
\affiliation{Department of Physics, Stevens Institute of Technology, Hoboken, NJ 07030, USA}
\affiliation{Department of Physics and Astronomy, University of Waterloo, Waterloo, Ontario, Canada, N2L 3G1}
\author{Dietrich Leibfried}
\affiliation{National Institute of Standards and Technology, 325 Broadway, Boulder, CO 80305, USA}
\author{Christian Sanner}
\affiliation{Department of Physics, Colorado State University, Fort Collins, CO 80523, USA}
\author{Igor Pikovski}\email{pikovski@stevens.edu}
\affiliation{Department of Physics, Stevens Institute of Technology, Hoboken, NJ 07030, USA}
\affiliation{Department of Physics, Stockholm University, Stockholm, Sweden}

 \begin{abstract}
Optical clocks based on atoms and ions probe relativistic effects with unprecedented sensitivity. They resolve time dilation due to atom motion or different positions in the gravitational potential through frequency shifts. However, all measurements of time dilation so far can be explained effectively as the result of dynamics with respect to a classical proper time parameter.
Here we show that atomic clocks can probe effects where a classical description of the proper time dynamics is insufficient as superpositions of proper time emerge.
We apply a Hamiltonian formalism to derive time dilation effects in harmonically trapped clock atoms and show how second-order Doppler shifts due to the vacuum energy, squeezing and quantum corrections to the dynamics arise. We also demonstrate that time-dilation-induced entanglement between motion and clock evolution can become observable in state-of-the-art clocks when the motion of the atoms is strongly squeezed, realizing proper time interferometry. 
Our results show that experiments with trapped ion clocks are within reach of probing relativistic evolution of clocks for which a quantum description of proper time becomes necessary.

\end{abstract}

\maketitle

\def\thefootnote{$\dagger$}\footnotetext{These authors contributed equally to this work}
\def\thefootnote{\arabic{footnote}}

\textit{Introduction}. 
Our understanding of the nature of time is continuously evolving. Time still plays different roles in the variety of physical theories that underpin modern physics. In quantum mechanics, it is usually treated as a fixed parameter, while in general relativity time becomes dynamical. But even in quantum mechanics governed by the Schr\"{o}dinger equation, we expect that time should change with velocity and position in a gravitational field, as dictated by relativity. Clocks that operate based on quantum principles evolve according to their proper time, and thus experience time dilation. This has been tested in many experiments 
\cite{choudoi:10.1126/science.1192720,Grotti2018,Takamoto2020,zheng2022differential, Bothwell2022}, with the first direct observation by Hafele and Keating \cite{hafele1972}. Such measurements rely on the precision of atomic clocks, which operate on fundamentally quantum mechanical principles \cite{ludlowRevModPhys.87.637}. Nevertheless to-date, even atomic clocks in experiments simply measure a \textit{fixed, classical} proper time parameter, probed by a quantum mechanical sensor \cite{takamoto2005optical,dubePhysRevA.87.023806,sanner2019lorentz,Bloom2014,campbell2017fermi}. Indeed, such experiments demonstrate that even the non-relativistic Schr\"{o}dinger-evolution of atomic clocks has to be governed by the proper time $\tau$. However, they do not probe other aspects of the interplay between quantum mechanical principles and relativistic principles.

Here we explore how genuine quantum-mechanical signatures of proper time evolution can be probed with atomic clocks. We use the Hamiltonian formalism introduced in \cite{zych2011quantum} to derive proper time signatures in clocks, such as the well-known second-order Doppler shift (SODS), or motional redshift, that stems from special relativistic time dilation \cite{choudoi:10.1126/science.1192720}. 
In this formalism it becomes clear that such effects can be generalized to include shifts from quantum mechanical effects, such as what we call here the ``vacuum-induced second order Doppler shift'' (vSODS). But while such effects stem from quantum motional states, they nevertheless can be reproduced by evolution with respect to a fixed semiclassical parameter $\langle \tau \rangle \approx t(1- \langle v^2 \rangle/(2c^2) )$.
Going beyond this description, we compute how time-dilation-induced entanglement \cite{zych2011quantum, pikovski2017time} and decoherence \cite{Pikovski2015} affect these systems. We show that the use of motional squeezed states allow for the first experimental demonstration of these effects with state-of-the-art technology, with an observable reduction of visibility, and an additional squeezing-induced frequency shift (sqSODS). Finally, we also derive what we term the ``quantum second-order Doppler shift'' (qSODS), which is an inherently quantum mechanical effect on the phase evolution of the clock, which can be amplified through control and measurement on the motional states. Our results show how the experimental exploration of a new regime of clock dynamics can be achieved, for which the quantization of the proper time evolution through $\hat{\tau} \equiv \tau(\hat{x},\hat{p})$ is necessary.

\textit{Formalism.}
All clocks evolve according to their local proper times $\tau$, including atomic clocks that evolve quantum mechanically \cite{Bothwell2022,Takamoto2020,Grotti2018,choudoi:10.1126/science.1192720,hafele1972,ludlowRevModPhys.87.637}. 
To describe this evolution in low-energy quantum systems, it is convenient to introduce a Hamiltonian formalism for clocks. Such a formalism essentially captures the relationship of proper time and coordinate time $t$, with respect to which the Hamiltonian generates the evolution \cite{zych2011quantum, pikovski2017time, Pikovski2015, orlando2017does, zych2016general, krause2017relativistic, sonnleitner2018mass, schwartz2019post, zychPhysRevD.99.104029, lahuertaPhysRevA.106.032803, hartong2024coupling}. Specifically, it was shown in \cite{zychPhysRevD.99.104029} that if a sufficiently small clock is described by its local Hamiltonian $H_\mrm{loc} = mc^2 + H_\mrm{c}$ that includes the rest-mass $m$ and clock evolution generator $H_\mrm{c}$, then the total evolution that describes both the clock and its motion is governed the Hamiltonian $H=\sqrt{(cp)^2 + H_\mrm{loc}^2}$. This is because the action capturing both the clock and its motion can be shown \cite{zychPhysRevD.99.104029} to be $S= -\int L_\mrm{loc} \mathrm{d}\tau$, where  $L_\mrm{loc}$ is the Lagrangian of the local clock degrees of freedom, whose Legendre transform is $H_\mrm{loc}$. Proper time evolution of clocks thus proceeds as if the rest mass $m$ is replaced by the total energy $mc^2 + H_\mrm{c}$ of the clock. The resulting quantized Hamiltonian for the joint clock and center-of-mass evolution was first considered in Ref. \cite{zych2011quantum} and applied to quantum interference and entanglement between clock and center-of-mass states in the gravitational field. Importantly, the quantum nature of the dynamics can yield new effects, such as time-dilation-induced entanglement \cite{zych2011quantum, Sinha_2011}, decoherence due to time dilation \cite{Pikovski2015, pikovski2017time}, interference of proper time evolutions \cite{Smith2020, grochowski2021quantum, Paczos2024quantumtimedilation, paczos2025witnessing}, corrections to atomic detector rates \cite{hu2024feasibility,wood2022quantized}, anomalous friction in atoms \cite{sonnleitner2018mass, sonnleitner2017will}, signatures in quantum networks \cite{barzel2024entanglement, borregaard2025testing, covey2025probing,fromonteil2025nonlocalmasssuperpositionsoptical}, and corrections to classical time dilation in boosted frames \cite{paigePhysRevLett.124.160602}. For our purposes, neglecting gravity and considering the clock in an external potential $\hat{V}$, we get $\hat{H} = mc^2 + \hat{p}^2/(2m) + \hat{V} + \hat{H}_\mrm{c}(1 -  \hat{p}^2/(2m^2c^2) )$, where we neglect terms beyond $O(c^{-2})$ and assume $\langle \h H_\mrm{c} \rangle \gg \langle \h V \rangle$, or in terms of clock and motional angular frequency, respectively: $\omega_\mrm{c} = 2 \pi \nu \gg \omega$. For a harmonic potential at trap frequency $\omega$, the Hamiltonian describing atomic and ion clocks with local evolution $\hat{H}_\mrm{c}$ and special relativistic time dilation, is thus

\begin{align}
    \hat{H} = \hat{H}_\mrm{c} + \hbar \omega \bigg( \hat{n} + \frac{1}{2} \bigg) - \frac{\hbar \omega}{2 m c^2} \hat{H}_\mrm{c} \hat{P}^2
    \label{eq:Htot}
\end{align}
with $\h P = \hat{p}/\sqrt{\hbar m \omega}=  i ( \h a^\dagger - \h a)/\sqrt{2}$ the momentum quadrature.

The Hamiltonian in Eq.\ \eqref{eq:Htot} can be interpreted in two equivalent ways: either as including the mass-energy equivalence (also sometimes called mass-defect in purely classical descriptions \cite{yudin2018mass}) due to internal energy $\smash{\hat{H}_\mrm{c}}$, or equivalently as incorporating relativistic proper time evolution to first order. Both pictures are simply different interpretations of the same Hamiltonian. Importantly, $\hat{H}$ only deviates from the more conventional nonrelativistic Hamiltonian through the last term which couples the clock degrees of freedom to the motion. This coupling has several consequences, and can entangle the clock with the motion, as we will discuss in the following.

We specifically model the clock as a two-level system with states $\ket{g}$ and $\ket{e}$ at energy zero and $\hbar \omega_\mrm{c}$ respectively, $\hat{H}_\mrm{c} =\hbar \omega_\mrm{c} |e\rangle\langle e|$, so that the entire rest energy when
the clock is in the ground state is captured by $mc^2$. It is useful to introduce the unitless operator

\begin{equation} \label{eq:epsilon}
    \hat{\varepsilon}_\mrm{c} = \varepsilon_\mrm{c} |e\rangle \langle e|\, ,
\end{equation}
with $\varepsilon_\mrm{c} = \hbar \omega_\mrm{c} / (mc^2) \ll 1$, the dimensionless ratio of the excited clock state energy to the rest energy. The corresponding dimensionless trap frequency parameter $\varepsilon_\mrm{m} = \hbar \omega / (mc^2)\ll 1$, is related to $\varepsilon_\mrm{c}$ by $\varepsilon_\mrm{m} \omega_\mrm{c} =  \varepsilon_\mrm{c} \omega$.
For the clock transition in $^{27}\textrm{Al}^+$ at 267~nm \cite{marshall2025high} and a trap frequency of $\omega=2\pi \times 20$~MHz, $\varepsilon_\mrm{m} = 3.29 \times 10^{-18}$  and $\varepsilon_\mrm{c} = 1.85 \times 10^{-10}$. With even lighter ions like $\textrm{B}^+$ and $\textrm{Li}^+$ and/or tighter confinement and higher transition frequencies, the $\varepsilon$ parameters will further increase. We note that $\textrm{Al}^+$-based clocks are typically operated in a two-ion configuration, which leads to a more complex motional mode structure. All $\varepsilon$ values considered in this paper are calculated for the simple case of a harmonically confined single ion.

\begin{figure}[t!]
    \centering
    \includegraphics[width=0.95\linewidth]{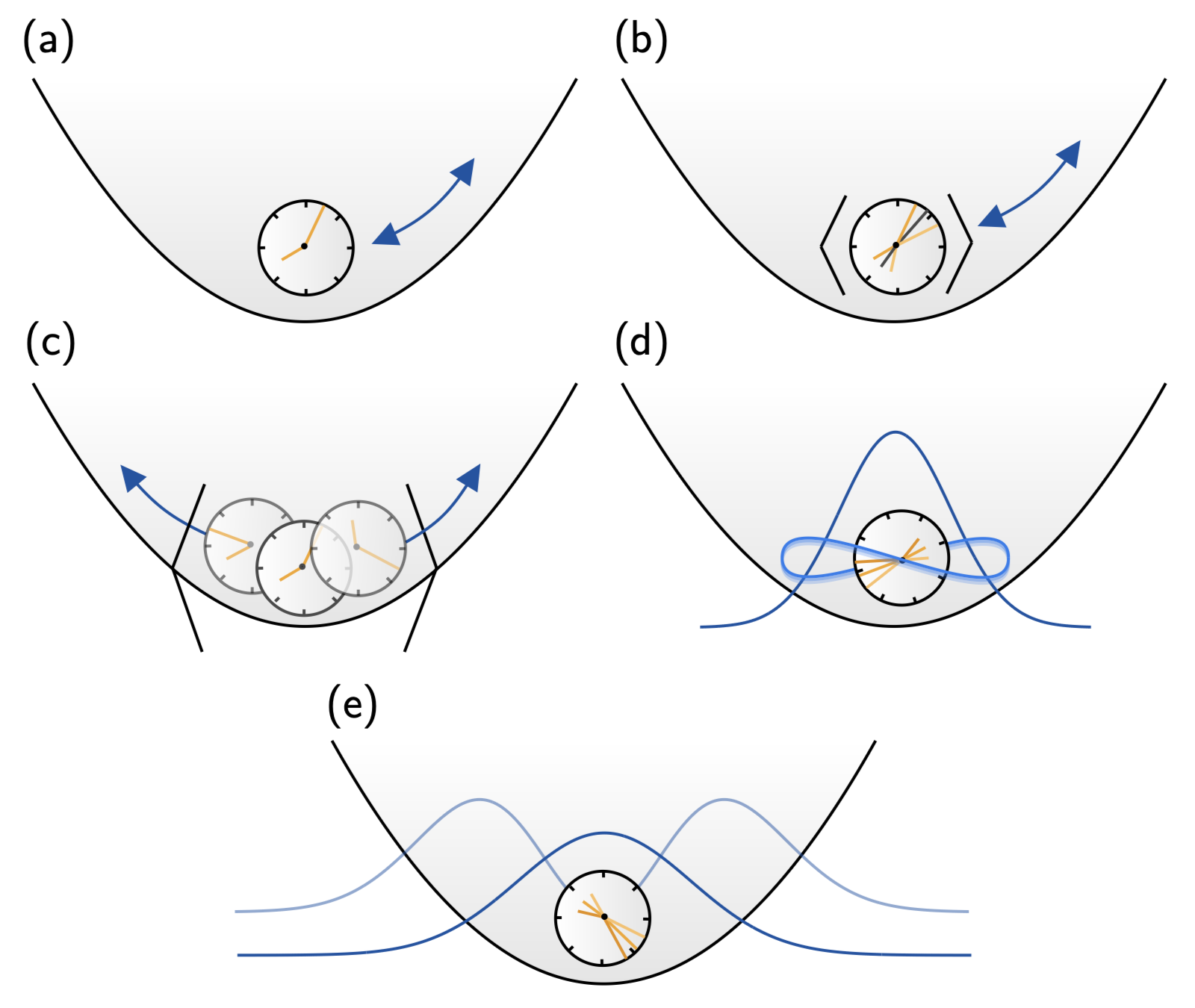}
    \caption{Illustration of classical, semiclassical, and quantum proper time dynamics of a trapped-ion atomic clock that we consider. In (a), a clock undergoes classical relativistic motion in a harmonic potential, and the clock evolution is $\hat{U}=e^{-i \hat{H}_\mrm{c} \tau/\hbar}$. In (b), the clock evolves with a single proper time that arises as an average with respect to a quantum motional state, $\hat{U}=e^{-i \hat{H}_\mrm{c} \langle\tau \rangle/\hbar}$. In (c), the proper time is averaged over the clock's classical mixture of motions $\hat{U}=\langle e^{-i \hat{H}_\mrm{c} \tau /\hbar}\rangle$. The full quantum evolution $\hat{U}=e^{-i \hat{H}_\mrm{c} \tau(\hat{x},\hat{p})/\hbar}$ leads to entanglement-induced loss of coherence (d) and the qSODS (e), in which the clock and spatial relativistic dynamics interfere.}
    \label{fig:1}
\end{figure}

The unitary evolution resulting from Eq.\ \eqref{eq:Htot} is derived in Appendix \ref{app:Unitary},  

\begin{align}\label{eq:Utot}
    \h U &= e^{-i \h H_\mrm{c} t / \hslash } \h S ( \h \zeta ) e^{- i \h \lambda \omega (\h n + 1/2 ) t }  \h S^\dagger ( \h \zeta ) 
\end{align}
where $\h S(\zeta) = e^{(\zeta/2)(\h a^2 - \h a^{\dagger 2} ) } 
$ is the squeezing operator and the arguments are operators themselves that act on the clock, $ \hat{\zeta} = - \ln ( 1 - \hat \varepsilon_\mrm{c} )/4 \simeq \h \varepsilon_\mrm{c}/4$ and $\hat{\lambda} = \sqrt{1 - \h \varepsilon_\mrm{c}} \simeq 1 - \h \varepsilon_\mrm{c}/2 $. Both operators depend on the clock Hamiltonian through Eq.\ \eqref{eq:epsilon} and since $\langle \hat{\zeta} \rangle \ll 1$ the approximation  
 $\hat{S}(\hat{\zeta})\simeq 1 + (\hat{\zeta}/2) ( \h a^2 - \h a^{\dagger 2} )$ is valid. The evolution operator in Eq.\ \eqref{eq:Utot} captures the well-known relativistic second-order Doppler shift in clocks, and additional quantum effects which can be probed with clocks.

We assume that the clock is prepared in the initial state $\ket{\psi(0)} = \left(\ket{g} + \ket{e} \right)\ket{\phi_\mrm{m}} /\sqrt{2} $, with motional state $\ket{\phi_\mrm{m}}$. The clock then evolves under Eq.\ \eqref{eq:Utot} into the state
\begin{align}\label{eq:state}
         | \psi(t) \rangle 
                = \frac{e^{-i\omega t \h n}}{\sqrt{2}} \left( | g \rangle + e^{-i\omega_\mrm{c}'(\h n)t} | e \rangle \right) | \phi_\mrm{m} \rangle + \h U_\mrm{cm} | e \rangle | \phi_\mrm{m} \rangle \, .
\end{align}
The first term shows that the clock frequency is altered to a new frequency  
\begin{align} \label{eq:nu}
    \hat{\omega}_\mrm{c}'(\hat n) &= \omega_\mrm{c} \bigg( 1 - \frac{\varepsilon_\mrm{m}(2\h n + 1)}{4} \bigg) \, .
\end{align}
It captures a frequency change that depends on the motional state through the operator $\hat{n}$, and is responsible for a variety of relativistic effects, as we show below. The second term in Eq. \eqref{eq:state} that depends on $\hat{U}_\mrm{cm}(\varepsilon_\mrm{c})$ (stemming from  $\h S ( \h \zeta )$, see 
Appendix \ref{app:Unitary})  captures a clock-motion coupling that induces additional changes in the motional state. It does not contribute to the standard SODS but it can result in the qSODS, as we will show further below.

\textit{SODS.}  
The exact clock evolution with respect to proper time 
follows from Eq.\ \eqref{eq:Utot}. If we neglect $\hat{U}_\mrm{cm}$ in Eq.\ \eqref{eq:state}, since it does not directly contribute to the frequency shift to first order,  we obtain the well-known SODS.  In this limit, the state evolves to

\begin{equation} \label{eq:vSODSstate}
    \ket{\psi(t)} = \frac{1}{\sqrt{2}} ( | g \rangle + e^{-i\omega_\mrm{c}'(\hat{n})t} | e \rangle ) | \phi_\mathrm{m} \rangle , 
\end{equation}
 and the off-diagonal elements of the clock, which capture its frequency and coherence, become

\begin{equation}\label{eq:offdiagonal}
2 \rho_\mrm{eg} = \left\langle e^{-i\omega_\mrm{c}'(\hat{n})t } \right\rangle .
\end{equation}
Here, the average is taken with respect to the initial motional state, which is arbitrary and can also be a mixed state.  Suppose that the motional state of the ion is thermal at temperature $T$, given by the state $\rho_\mrm{th}$.The clock experiences a frequency shift and drop in interference contrast purely from the proper time evolution, however the loss of coherence scales as $1-(\varepsilon_\mrm{c} \omega t\bar{n})^{2}$ and can be neglected in currently realizable experiments since it is of second order in $\varepsilon_\mrm{c}$ (see Appendix \ref{A:thermal}). 
The frequency shift, to first order in $\varepsilon_\mrm{c}\omega t\bar{n}$, is

\begin{equation} \label{eq:SODS}
    \frac{\Delta\nu_\mrm{SODS}}{\nu} 
    =  - \frac{\hbar \omega (2\bar n+1)}{4mc^2}  \simeq - \frac{k_B T}{2 mc^2} = - \frac{\langle v^2 \rangle_{\textrm{th}}}{2 c^2}
\end{equation}
where $2 \pi \Delta\nu_\mrm{SODS}= \omega_\mrm{c}' - \omega_\mrm{c}$ and $\langle v^2 \rangle_\mrm{th}$ is the mean square velocity of the thermal distribution, and we used $\bar{n} \approx k_B T/\hbar \omega$. The result thus reproduces the SODS (here computed in 1-dimension),as accounted for in many experiments \cite{ludlowRevModPhys.87.637}, and shown in Fig. \ref{fig:1}b.This simple example shows the strength of the formalism employed here; the SODS simply follows from the evolution due to the Hamiltonian Eq.\ \eqref{eq:Htot}. In fact, the general result for the thermal state as described in the Appendix \ref{A:thermal}, where the full average in Eq.\ \eqref{eq:offdiagonal} is relevant, shows that the SODS can also take other forms which can become observable at low temperatures. This is illustrated in Fig. \ref{fig:1}c.

\textit{vSODS.} We now consider an ion that is cooled to the motional ground state, for which $\ket{\phi_\mrm{m}} = \ket{0}$. We again neglect $\hat{U}_\mrm{cm}$ in Eq.\ \eqref{eq:state}. Despite being in the ground state, the dynamics yields a shift in the clock frequency according to Eq.\ \eqref{eq:SODS} with $\bar{n}=0$, consistent with previous calculations \cite{haustein2019massenergyequivalenceharmonicallytrapped, lahuertaPhysRevA.106.032803}

\begin{equation} \label{eq:vSODS}
    \frac{\Delta\nu_\mrm{vSODS}}{\nu} = -\frac{\varepsilon_\mrm{m}}{4} = - \frac{\hbar \omega}{4 m c^2}  \, .
\end{equation}
We stress that this effect arises despite cooling to the absolute ground state of the ion. This is because of vacuum fluctuations, or more specifically because proper time is velocity-dependent and the ground state is not a momentum eigenstate, i.e., the finite spatial extent of the wavefunction implies a finite spread of momentum around zero. The vSODS is thus an example of an effect where proper time evolution is not due to a classical well-defined state, and neither stemming from a classical mixture. It arises because of the clock picking up vacuum contributions through the factor $\hat{\varepsilon}_\mrm{c}\omega t/4$ in Eq.\ (\ref{eq:Utot}), thus being an observable manifestation of the quantum vacuum.

There is currently great interest in observing effects from proper time that stem from inherently quantum mechanical states of motion. This has been considered for the gravitational case \cite{zych2011quantum, pikovski2017time, Khandelwal2020universalquantum,Paczos2024quantumtimedilation}, and for special relativistic time dilation \cite{Smith2020,grochowski2021quantum,hu2024feasibility}. The vSODS discussed above is one such example. However, there are ambiguities when measuring the frequency shift. If entanglement was to be measured as proposed in \cite{zych2011quantum} and adapted to our case below, then the outcome would be inconsistent with a classical picture of proper time, i.e.\ to explain entanglement it is really necessary to quantize also the experienced proper time as $\hat{\tau} = \tau(\hat{x},\hat{p})$. In contrast, signatures on the frequency, such as the vSODS, cannot distinguish the purely quantum-mechanical description from a semiclassical model in which we replace $\hat{\tau}$ by just $\langle \tau\rangle $. This is illustrated in Fig. \ref{fig:1}b, where now the expectation value is with respect to the vacuum state. We can generalize the result: To order $O ( \varepsilon_\mrm{c} )$, the frequency shift in Eq.\ \eqref{eq:offdiagonal} becomes $\langle e^{-i\omega_\mrm{c}'(\hat{n})t } \rangle \simeq e^{-i\omega_\mrm{c}'( \langle \hat{n} \rangle)t }$, where $\omega_\mrm{c}'( \langle \hat{n} \rangle) = \omega_\mrm{c}(1-\varepsilon_\mrm{m}m \langle v^2 \rangle/(\hbar \omega))$. The clock thus evolves as if simply due to the classically averaged proper time, $2 \rho_\mrm{eg} \simeq e^{i \omega_\mrm{c} \langle \tau \rangle} = \exp (i \omega_\mrm{c} t (1- \langle v^2 \rangle/(2c^2) )  )$. This holds for any motional state and thus can generically reproduce the clock's frequency shift semiclassically. In particular, it reproduces both the SODS in Eq. \eqref{eq:SODS} and the vSODS in Eq. \eqref{eq:vSODS}. Even beyond the perturbative expansion, the frequency shift can always be captured by Eq.\ \eqref{eq:offdiagonal}, which admits a classical interpretation. Given an initial classical energy $E$, averaged over the possible initial energies of the motion, the clock frequency shifts to $\omega_\mrm{c}'= \omega_\mrm{c}(1- E/(mc^2) )$.

\begin{figure}[t!]
    \centering
    \includegraphics[width=\linewidth]{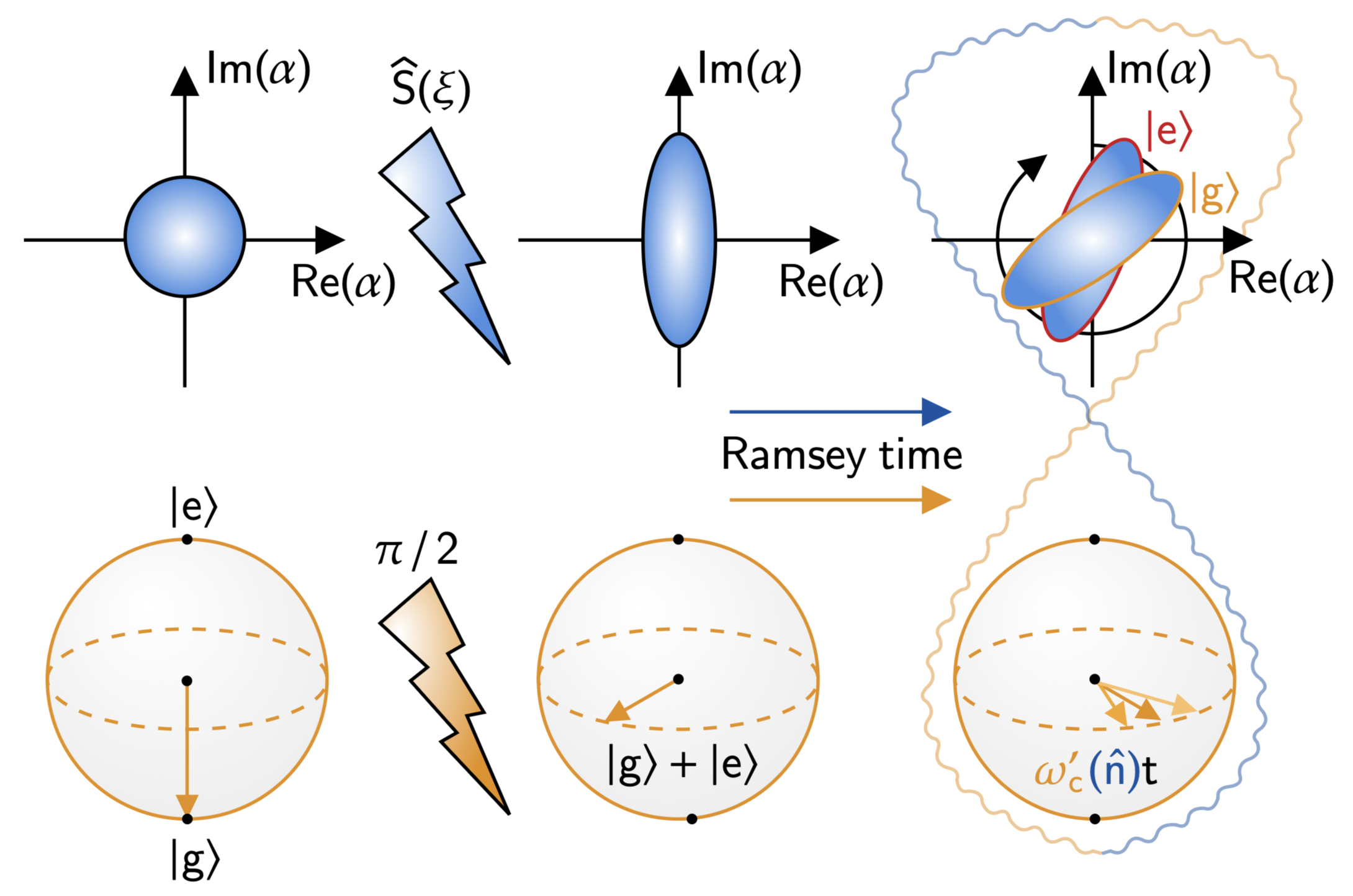}
    \caption{Illustration of time-dilation-induced entanglement between clock and motional degrees of freedom, and how it can be observed using trapped atomic clocks with squeezing of motional states. The protocol proceeds from left to right. The top row shows the motional states in a phase-space representation, where a squeezed state is prepared which then evolves at different frequencies in superposition, depending on the internal clock-states. The bottom row shows the same sequence from the perspective of the clock degrees of freedom represented on a Bloch sphere, where a Ramsey sequence results in a superposition of different time evolutions of the clocks due to the different motional energies. The entanglement between motion and clock (last column) causes a reduction in visibility of the clock, Eq.\ \eqref{eq:visibility}, which can be measured with current state-of-the-art ion clock systems.}
    \label{fig:2}
\end{figure}

\textit{Time-dilation-induced entanglement.} A key feature of the genuine quantum evolution in Eq. \eqref{eq:Utot} is that entanglement can be generated between clock and motional states through the proper time evolution. This can happen even if the clock-dependent squeezing operators, which result in $\hat{U}_{\textrm{cm}}$, are neglected. The entanglement arises through the motion-dependent frequency shift in Eq.\ \eqref{eq:nu}, illustrated in Fig. \ref{fig:1}d. It was first predicted in Ref.\ \cite{zych2011quantum} with a focus on gravitational time dilation, and has been generalized to many different systems and situations \cite{zych2012general,pikovski2017time,bushev2016single,loriani2019interference,castro2020quantum,roura2020gravitational,roura2021measuring,barzel2024entanglement,meltzer2024atomic,borregaard2025testing,covey2025probing}. But the predicted time-dilation-induced entanglement of clock states to motional states has so far not been observed (other than in an experiment that simulated time dilation \cite{margalit2015self}). We now show how it can be realistically measured in ion clocks, summarized in Fig. \ref{fig:2}. To this end, we focus on the resulting loss of coherence as also in  \cite{zych2011quantum,Pikovski2015,pikovski2017time}, which is captured by the dimensionless interferometric visibility, in our case $V=2|\rho_{\textrm{eg}}|=| \langle e^{-i\omega_\mrm{c}'(\hat{n})t } \rangle | $. Instead of using the motional ground state, we now consider an initial squeezed vacuum state $\ket{\xi} = \hat{S}(\xi)\ket{0}$, with $|\xi|=r$. The off-diagonal elements of the clock now become
\begin{align}\label{eq:squeeze}
    2 \rho_\mrm{eg} = 
     \frac{ e^{-i \omega_\mrm{c} t ( 1 - \varepsilon_\mrm{m}/4 ) } }{\sqrt{ \cosh^2(r) - e^{i\omega_\mrm{c} \varepsilon_\mrm{m}t} \sinh^2(r) } }
\end{align}
where we used $e^{-i\omega_\mrm{c}'(\hat n )t} \h S ( \xi ) e^{i\omega_\mrm{c}'(\hat n)t} = \h S (\xi e^{i\omega_\mrm{c} \varepsilon_\mrm{m}t/2})$ and the overlap of squeezed states $\langle r | r e^{i\varphi} \rangle = ( \cosh^2(r) - e^{i\varphi} \sinh^2(r) )^{-1/2}$. For no initial squeezing, the result reduces to the vSODS described before. However, the squeezing now introduces an additional frequency shift
\begin{equation}\label{eq:sqSODS}
    \frac{\Delta \nu_\mrm{sqSODS}}{\nu} \simeq - \frac{\varepsilon_\mrm{m}}{4}\mrm{cosh}(2r) 
\end{equation}
and $V = \left(\mrm{cosh}^2(2r) \, \mrm{sin}^2 \left( \frac{\varepsilon_\mrm{m}  \omega_\mrm{c} t}{2} \right) + \mrm{cos}^2 \left( \frac{\varepsilon_\mrm{m} \omega_\mrm{c}  t}{2} \right)\right)^{-1/4}$ for the visibility, which is approximately
\begin{equation}\label{eq:visibility}
\begin{split}
    V  \simeq 1 - \frac{(\varepsilon_\mrm{m}  \omega_\mrm{c} t)^2}{16} \mrm{sinh}^2(2r)  \, .
\end{split}
\end{equation}
The visibility loss arises due to the entanglement generated between motional and clock states through the proper time evolution in Eq.\ \eqref{eq:Htot}, illustrated in the last column of Fig. \ref{fig:2}. It is a witness of the inherent quantum proper time evolution \cite{zych2011quantum}. Our result suggests that this may be observable in a state-of-the art experiment.  The $^{27}\textrm{Al}^+$-clock \cite{marshall2025high} gives $\varepsilon_\mrm{m} = 3.3 \times 10^{-18}$. A $20~$MHz trap with excellent clock and motional coherence to permit a free evolution of duration $t \simeq 1~$s and further assuming that squeezed states with $r=2.26$ can be prepared  (such states have already been observed in an ion trap \cite{burd2019quantum}) would result in a reduction of visibility in Eq.\ \eqref{eq:visibility} to $V\simeq 0.93$.

The resulting frequency shift would also be observable, $\Delta\nu_\mrm{sqSODS}/\nu \simeq 3.8 \times 10^{-17}$, but the visibility loss would be of greater interest as it is the hallmark of the underlying quantum dynamics and entanglement. Thus our protocol allows the observation of the time-dilation-induced entanglement, which goes beyond the standard classical treatment of proper time dynamics. In fact, it is in principle also possible to directly verify this entanglement, by joint measurements both on the motional and clock states, without relying on the visibility as a witness. 

\def\arraystretch{1.5}
\begin{table*}[t]
\centering
\begin{tabular}{|c|c|c|c|} 
\hline
 & \rule{0pt}{1ex} Initial Motional State:\,  $\ket{\phi_\mrm{m}}$ 
 & Fractional Frequency Shift: \, $\frac{\Delta\nu}{\nu}$ 
 & Visibility:\quad $V$ \\[0ex] \hline

SODS
& \rule{0pt}{1ex} $\rho_{\mrm{th}}$
& $ - \frac{\langle v^2\rangle_\mrm{th}}{2c^2}$ 
& $ 1-\frac{1}{8}(\varepsilon_\mrm{c} \omega t)^2 \bar{n}(\bar{n}+1)$
\\[0ex] \hline

vSODS
& \rule{0pt}{1ex} $\ket{0}$
& $ -\frac{\varepsilon_\mrm{m}}{4}$ 
& $ 1-\frac{\varepsilon_\mrm{c}^2}{16}\sin^2(\omega t)$
\\[0ex] \hline

sqSODS
& \rule{0pt}{1ex} $\hat{S}(\xi)\ket{0}$
& $ -\frac{\varepsilon_\mrm{m}}{4}\cosh(2r)$ 
& $ 1-\frac{(\varepsilon_\mrm{c} \omega t)^2}{16}\sinh^2(2r)$
\\[0ex] \hline

qSODS
& \rule{0pt}{1ex} $\ket{0}$
& $\begin{array}{c}
-\frac{\varepsilon_\mrm{m}}{4}-\frac{\varepsilon_\mrm{m}\varepsilon_\mrm{c}}{16} \left(1 - \textrm{sinc}(2\omega t) \right) \\ \text{with control:} \quad  
-\frac{\beta\varepsilon_\mrm{c}}{\sqrt{8}\nu t(2+\beta^2)} 
\end{array}$
& $1- \frac{\varepsilon_\mrm{c}^2}{16}\sin^2(\omega t)$
\\[0ex] \hline
\end{tabular}
\caption{
Summary of the manifestations of the second-order Doppler shifts that arise from the relativistic dynamics given by Eq. \eqref{eq:Htot}, to lowest order in $\varepsilon_\mrm{c}$ and $\varepsilon_\mrm{m}$. The proper time evolution of the clocks results in relativistic frequency shifts, and in a drop in visibility due to the clock-motion-coupling. Thermal motional states give rise to the standard SODS, while the quantum ground state causes the vacuum-SODS. Going beyond effects that can be described via a classical parameter $\langle \tau \rangle $, preparing a motional squeezed state allows one to resolve the entanglement-induced visibility drop for the sqSODS with near-future ion clocks (see also Fig. \ref{fig:2}). The qSODS is an additional quantum contribution from the action of $U_\mrm{cm}$ in Eq. \eqref{eq:state}.}
\label{table:summary}
\end{table*}

\textit{qSODS.} 
We now consider the full evolution, Eq.\ \eqref{eq:Utot}, which includes contributions from  $\hat{S}(\hat{\zeta})$ that lead to $\hat{U}_\mrm{cm}$ in Eq. \eqref{eq:state}. This squeezing is a consequence of the quantum relativistic evolution of the clock in the trap, not from the initial state as above. The unitary $\hat{U}_\mrm{cm}$ depends on $\hat \zeta \simeq \hat \varepsilon_\mrm{c}/4$, which we have neglected in our calculations thus far. This contributes an additional shift in the clock, which we call the quantum second-order Doppler shift (qSODS), illustrated in Fig. \ref{fig:1}e.
 
Let us assume the ion is initially cooled to the ground state. The evolved state to leading order in $\varepsilon_\mrm{c} $ is
\begin{align}\label{eq:qSODSstate}
    | \psi (t) \rangle &\simeq | \psi_\mrm{c,0}(t) \rangle - \frac{1}{8} \varepsilon_\mrm{c}h(t)e^{-i\omega_\mrm{c}'(0)t} | e \rangle | 2 \rangle  
\end{align}
where $| \psi_\mrm{c,0}(t) \rangle$ captures the vSODS contribution to the clock evolution, Eq.\ (\ref{eq:vSODSstate}), and $h(t) = 1 - e^{-2i\omega t (1 - \varepsilon_\mrm{c}/4 )} $. 
In regular clock measurements, the additional contributions in Eq.\ \eqref{eq:qSODSstate} will only cause visibility oscillations of order $\varepsilon_\mrm{c}^2$, after tracing over the motional state. The exact reduced clock state is computed in Appendix \ref{app:HigherOrderState}: the off-diagonal elements are given by $2 \rho_\mrm{eg} = e^{-i \omega_\mrm{c} t (1- \varepsilon_\mrm{m}/4 )} \left(\mrm{cosh}^2(\varepsilon_\mrm{c}/4) - e^{i \omega t (2-\varepsilon_\mrm{c}) } \mrm{sinh}^2(\varepsilon_\mrm{c}/4)\right)^{-1/2}$. This yields an  additional phase shift of $\varphi \simeq \varepsilon_\mrm{c}^2(2 \omega t - \sin(2\omega t))/32$ and a visibility drop of $V \simeq 1 - (\varepsilon_\mrm{c} \sin(\omega t)/4)^2$, both too small to observe with currently feasible experiments.

Instead of direct clock measurements, one can design protocols to reveal the additional contributions to linear order in $\varepsilon_\mrm{c}$ that are not averaged out, see the Supplemental Material \cite{SM}. In brief, one can project the motional states on $( | 0 \rangle + | 2 \rangle ) /\sqrt{2}$ rather than tracing them out, which results in a signal on the clock that is linear in $\varepsilon_\mrm{c}$. To also convert the highly oscillatory signal into a fixed phase offset, one needs to apply clock-state-dependent operations on the motional state before the measurement. This can, for example, be a clock-dependent displacement or linear drive. Such operations result in a phase offset on the clock on the order of $\phi_\mrm{qSODS} \simeq \varepsilon_\mrm{c}$, which stems from the $\hat{S}(\hat{\varepsilon}_\mrm{c}/4)$ contribution to the relativistic dynamics. It represents an additional quantum mechanical effect as it cannot be captured by the semiclassical evolution with respect to $\langle \tau \rangle$. However, while this phase offset is now linear in $\varepsilon_\mrm{c}$, even for the $^{27}\textrm{Al}^+$-clock it is of order $\phi_\mrm{qSODS} \simeq 10^{-10}$~rad, too small to observe with current and near-future capabilities.

\textit{Conclusions.} 
In summary,we showed how experiments with harmonically confined clock atoms can probe quantum features of the relativistic proper time dynamics, which have so far not been observed. We used the Hamiltonian formulation of relativistic dynamics of composite systems for a convenient description of internal and motional relativistic effects in atomic clocks due to special relativity. We showed how well-known effects are captured in this framework, and derived additional effects due to the quantization of clocks and motion, as summarized in table \ref{table:summary}. The vacuum-induced second order Doppler shift, which stems from the vacuum energy contributions of the trapping potential, leads to a shift of $\sim 5 \times 10^{-19}$ in a MHz trap, and thus may be observable. We also showed that the entanglement between internal clock states and external motional states generated through time dilation may become observable with $^{27}\textrm{Al}^+$-clocks that also feature state-of-the-art capabilities in squeezing and long coherence times.  The quantum dynamics also cause additional contributions which can give rise to the qSODS, which however is currently unobservable. Further improvements are possible, for example with higher trap frequencies or the development of a $^{10}\textrm{B}^+$-ion clock at $\nu = 1119$ THz \cite{wineland2002quantum,arnold2015prospects}, which due to its lower rest mass would result in a visibility drop to even $V\sim0.76$ for all else equal. Overall, our results show that the seemingly simple proper time evolution of clocks offers new phenomena and unique quantum features that can be probed with trapped ion-clocks.

\textit{Acknowledgements}. We thank Konstantin Beyer, Tara Fortier, David Hume, Nils Huntemann, Myungshik Kim, Shimon Kolkowitz, Alexander R.H.\ Smith, Jun Ye, and Magdalena Zych for helpful discussions during the development of this work, and David Hume also for a careful reading of the manuscript. This work was supported by the National Science Foundation under Grant No.\ 2239498 and Grant No.\ 2409166, by NASA under Grant No.\ 80NSSC25K7051 and by the ``Table-top experiments for fundamental physics'' program 
sponsored by the Alfred P. Sloan Foundation, Gordon and Betty Moore Foundation, 
John Templeton Foundation, and Simons Foundation
under Grant No.\ G-2023-21102. J.F.\ acknowledges funding from the Natural Sciences and Engineering Research Council through a Banting Postdoctoral Fellowship. G.S. acknowledges funding from the Novartis Science Scholarship.


%

\pagebreak

\pagebreak
\appendix

\section{Derivation of the Total Unitary Operator for Proper Time Evolution of Clocks and Motion}\label{app:Unitary}

The Hamiltonian in Eq.\ (\ref{eq:Htot}) generates the time evolution given by the unitary operator,
\begin{align}
    \h U &= e^{-i \h H_\mrm{c}t/\hslash} e^{-i \omega t ( \h P^2 + \h X^2 )/2 + i\omega t \h \varepsilon_\mrm{c}  \h P^2 / 2 } 
    \label{eqa1unitary}
\end{align}
where $\h P = i ( \h a^\dagger - \h a ) /\sqrt{2}$ and $\h X = ( \h a + \h a^\dagger ) /\sqrt{2}$ are the momentum and position quadratures of the motion, and $\h \varepsilon_\mrm{c} = \h H_\mrm{c}/(mc^2)$ as previously defined. We want to decouple this exponential into a product of operators to easily compute its effect of the clocks and motion. This can be achieved as follows: Acting with squeezing operators on both sides gives
 
\begin{align}
    & \h S^\dagger ( \zeta ) \h U \h S ( \zeta ) 
    \nonumber
    \vt \\
    &= e^{-i \h H_\mrm{c} t /\hslash } e^{- i \omega t ( \h P^2 e^{2\zeta} + \h X^2 e^{-2\zeta} )/2 + i \omega t \h \varepsilon_\mrm{c} \h P^2 e^{2\zeta}/2 } \, .
    \label{eqa2unitary}
\end{align}
One can choose $\zeta$ such that the exponential is only a rotation in the motional states by setting the coefficients of $\h P^2, \h X^2$ in Eq.\ (\ref{eqa2unitary}) to be equal. This gives the condition $\zeta \equiv \zeta ( \h H_\mrm{c} ) = - \ln ( 1 - \h \varepsilon_\mrm{c} )/4$, where it is now an operator acting on the clock states, and allows us to write Eq.\ (\ref{eqa2unitary}) as
\begin{align}
    \h S^\dagger ( \h \zeta ) \h U \h S( \h \zeta ) &= e^{-i \h H_\mrm{c}t/\hslash} e^{- i \hat{\lambda} \omega ( \h n + 1/2 ) t } 
    \vt 
    \label{eqa4unitary}
\end{align}
having additionally defined $\h \lambda \equiv \lambda ( \h H_\mrm{c} ) = \sqrt{1 - \h \varepsilon_\mrm{c}}$. Moving the operators $\h S^\dagger ( \h \zeta )$ and $\h S (\h \zeta )$ to the right-hand side gives 
\begin{align}
    \h U &= e^{-i \h H_\mrm{c} t / \hslash } \h S ( \h \zeta ) e^{- i \hat{\lambda} \omega (\h n + 1/2 ) t }  \h S^\dagger ( \h \zeta )  \, ,
\end{align}
which is Eq. (\ref{eq:Utot}) in the main text. We note that the clock Hamiltonian is arbitrary up to this point, which reflects the universality of time dilation. For our purposes, we use a two-level system as the clock.

The unitary evolution is dominated by the rotation with respect to $\lambda$, and the squeezing operations are subdominant. This can be seen by swapping the rotation and squeezing operations, which yields:
\begin{equation} \label{eq:Utot1}
    \h U = e^{-i \h H_\mrm{c} t / \hslash } \h S ( \h \zeta ) \h S^\dagger ( \h \zeta e^{ 2 i \hat{\lambda} \omega t }  )  e^{- i \hat{\lambda} \omega (\h n + 1/2 ) t }   \, .
\end{equation}
The squeezing operations thus effectively cancel, except for the additional rotation through $2 \hat{\lambda} \omega t$ that depends on the clock state. This can be better seen by combining the two squeezers, which to order $O(\zeta^2)$ is given by $\h S ( \h \zeta ) \h S^\dagger ( \h \zeta e^{ 2 i \hat{\lambda} \omega t }  ) = \h S \left( \h \zeta ( 1 - e^{ 2 i \hat{\lambda} \omega t  })  \right) e^{ i  \hat{\zeta}^2\sin(2\omega t) (\h n + 1/2 ) } +O(\zeta^3)$.
It shows that $\zeta$ contributes only at second order to the phase-shift, while also affecting the motional state through the squeezing operation, which gives rise to the qSODS in the main text.

To lowest order,  we can express its effect perturbatively, using the approximation $\varepsilon_\mrm{c} \ll 1$ as in the main text. We can now write the evolution operator as
\begin{align}
    \h U &= e^{-i \h\lambda \omega t ( \h n + 1/2 ) } + \frac{\h \varepsilon_\mrm{c}}{8} e^{-i \h\lambda \omega t ( \h n + 1/2 ) } \big( \h a^{\dagger 2} - \h a^2 \big) 
    \nonumber \\
    & + \frac{\h \varepsilon_\mrm{c}}{8} \big( \h a^2 - \h a^{\dagger 2} \big) e^{-i\h\lambda \omega t ( \h n + 1/2 ) } + O ( \h \varepsilon_\mrm{c}^2 ) 
    \label{eqaA6unitary}
\end{align}
The first term in Eq.\ (\ref{eqa1unitary}) gives rise to the usual SODS frequency shift in atomic clocks, and depends on the motional state through the operator $\h n$. The second and third terms constitute $\h U_\mrm{cm}$ as denoted in Eq.\ (\ref{eq:state}) and capture the squeezing of the motional state (i.e.\ energy transitions $n \to n + 2, n - 2$) to leading order in $\h \varepsilon_\mrm{c}$. The signatures on the clock to higher order and for a general initial state are computed in the next section.

\section{State Evolution for Initial Thermal State} \label{A:thermal}

We start with Eq.\ (\ref{eq:state}), neglecting contributions due to $\hat U_\mrm{cm}(\h \varepsilon_\mrm{c})$. We assume that the clock is initially in a thermal state $\rho_\mrm{th} = \sum_k ( 1 + \bar n)^{-1} ( \bar n/(1+\bar n))^k | k \rangle\langle k |$. Tracing over the motional degrees of freedom, gives the reduced state of the clock,

\begin{align}
    \rho_\mrm{c}(t) &= \sum_k \frac{\bar n^k}{(1+\bar n)^{k+1}} | \psi_{\mrm{c},k}(t) \rangle\langle \psi_{\mrm{c},k}(t) | 
\end{align}
where $| \psi_{\mrm{c},k}(t) \rangle = ( | g \rangle + e^{-i\omega_\mrm{c}'(k)t} | e \rangle ) / \sqrt{2}$. After performing the sum over $k$, the off-diagonal element in polar form is given by
\begin{align}
    2\rho_\mrm{eg} 
        &= \frac{e^{-i ( \omega_\mrm{c} t - \arctan( \tan(\varepsilon)(2\bar{n}+1) ))}}{\sqrt{\cos^2(\varepsilon) + \sin^2(\varepsilon)(2\bar n+1)^2}} 
\end{align}
having here defined $\varepsilon= \varepsilon_\mrm{c}\omega t/4$. We can read off from this expression the frequency shift $\Delta \omega_\mrm{c}$ and the visibility $2 |\rho_\mrm{eg}|$. In the regime $\varepsilon\bar{n} \ll 1$, we obtain to $O ( \varepsilon\bar n)$ the usual SODS as in the main text: 
\begin{equation} 
    2\rho_\mrm{eg} \simeq  \frac{e^{-i( \omega_\mrm{c} t- \varepsilon(2\bar{n}+1)) } }{\sqrt{1-\varepsilon^2+ \varepsilon^2(2\bar{n}+1)^2}} 
    \label{eqthermalstate}
\end{equation}
The visibility is suppressed at $O ( \varepsilon^2\bar{n}^2)$, but the frequency shift is $O( \varepsilon\bar{n})$, given in the main text in Eq.\ (\ref{eq:SODS}).

Now consider a high temperature limit where $\varepsilon\bar{n} \gtrsim 1$, but still $\varepsilon\ll 1$. This results in 
\begin{equation}
    2\rho_\mrm{eg} \simeq  \frac{e^{-i( \omega_\mrm{c} t- i\arctan(2\varepsilon \bar{n}))}}{\sqrt{1+4\varepsilon^2\bar{n}^2}} 
\end{equation}

The visibility drops to zero and no frequency shift can be observed as the temperature increases. However for intermediate temperatures or large trap frequencies such that $\varepsilon \bar{n} \sim 1$, one can observe a frequency shift that deviates from the usual SODS, Eq. \eqref{eqthermalstate}. In this regime, the effect of thermal motion is to not simply cause a single time dilation due to $v^2_\mrm{th}$, but rather an averaging of time dilations over different samples from the thermal velocity distribution. Of course observing such a shift would be challenging for a noisy system, however it can be attained at moderately low $\bar{n}$ if $\varepsilon = \hbar \omega \omega_\mrm{c} t/(4mc^2) \simeq 1/\bar{n}$.

\section{Frequency shift and visibility loss in clocks}\label{app:HigherOrderState}

We derive the reduced state of the clock which captures its phase evolution and also loss of visibility due to the entangling evolution \eqref{eq:Utot}. 
Assuming that we have an arbitrary initial motional state $\rho_\mrm{m}$ and the initial 
state of a two-level clock $\ket{\psi_\mrm{c}}= \left(\ket{g}+\ket{e}\right) /\sqrt{2}$, we begin with the joint initial state state
\begin{equation} \label{eq:initialRho}
    \rho_0= \ket{\psi_\mrm{c}}\bra{\psi_\mrm{c}}\otimes \rho_\mrm{m} \, .
\end{equation}
To evolve this state we will use the spectral decomposition in the clock states of our unitary \eqref{eq:Utot}, 
\begin{align}
    & \hat{U}= \hat{U}_\mrm{g}\ket{g}\bra{g}+\hat{U}_\mrm{e}\ket{e}\bra{e}
\end{align}
where $\hat{U}_\mrm{g}$ and $\hat{U}_\mrm{e}$ act only on the motional state and are given by
\begin{align}
    & \hat{U}_\mrm{g}=e^{-i\omega t (\hat{n}+1/2)}
    \vt 
    \label{eqB3}
    \\
    & \hat{U}_\mrm{e}=e^{-i\omega_\mrm{c} t}S({\zeta})e^{-i{\lambda}\omega t (\hat{n}+1/2)}S^\dag({\zeta})
    \vt 
    \label{eqB4}
\end{align}
with $\lambda = \sqrt{1-\varepsilon_c}$. Evolving the initial state, Eq.\ \eqref{eq:initialRho}, gives  
\begin{equation}
    \rho(t)= \frac{1}{2}\left(\hat{U}_\mrm{g}\ket{g}+\hat{U}_\mrm{e}\ket{e}\right) \rho_\mrm{m} \left(\bra{g}\hat{U}_\mrm{g}^{\dagger}+\bra{e} \hat{U}_\mrm{e}^{\dagger}\right)
\end{equation}
Now we can find the off-diagonal elements in the state of the clock and take the partial trace with respect to the motional state: 
\begin{align}
        2\rho_\mrm{eg} &= 2\textrm{Tr}_\mrm{m}\Big( \langle g | \rho(t) | e \rangle \Big) = \textrm{Tr}_\mrm{m}\left( \hat{U}_\mrm{g}\rho_\mrm{m}\hat{U}^\dagger_\mrm{e} \right) 
        \nonumber 
        \vt \\
        &= \left\langle\hat{U}^\dagger_\mrm{e}\hat{U}_\mrm{g}\right\rangle 
        \vt 
\end{align}
where the average is taken with respect to the arbitrary initial motional state $\rho_\mrm{m}$. This expression captures both the frequency shift and the visibility of the clock.
 Now inserting our definitions from Eqs.\ (\ref{eqB3}) and (\ref{eqB4}), 
\begin{equation}
\begin{split}
       &2\rho_\mrm{eg}= e^{i\omega_\mrm{c} t}\biggl\langle S({\zeta})e^{i{\lambda}\omega t (\hat{n}+1/2)}S^\dagger({\zeta}) e^{-i\omega t (\hat{n}+1/2)}\biggr\rangle\\
       &=e^{i\omega_\mrm{c} t+\frac{i\omega t}{2}(\lambda-1)}\biggl\langle S({\zeta})S^\dagger({\zeta}e^{2i{\lambda}\omega t })e^{i\omega t\hat{n}(\lambda-1)}\biggr\rangle 
\end{split}
\end{equation}
Note that if we neglect the squeezing operations and work to leading order in $\varepsilon_\mrm{c}$, we have exactly the result in Eq.\ \eqref{eq:offdiagonal} in the main text.

Now we can consider a case when $\rho_\mrm{m}= \ket{0}\bra{0}$, which is equivalent to Eq.\ \eqref{eq:vSODS} but with the inclusion of the additional squeezing operations that arise from the proper time evolution, and which we denoted by $\hat{U}_\mrm{cm}$ in the main text. Recall that $\zeta \in \mathbb{R}$, thus:
\begin{equation} \label{eq:FullStateClock}
\begin{split}
    2\rho_\mrm{eg}&= e^{i\omega_\mrm{c} t+\frac{i\omega t}{2}(\lambda-1)}\bra{0} S({\zeta})S^{\dagger}({\zeta}e^{2i{\lambda}\omega t })\ket{0}
    \vphantom{\bigg)} 
    \\
    &=  e^{i\omega_\mrm{c} t+\frac{i\omega t}{2}(\lambda-1)}\bra{{-{\zeta}}}\ket{-\zeta e^{2i{\lambda}\omega t }} \\
    &= \frac{e^{i\omega_\mrm{c} t +\frac{i\omega t}{2} (\lambda-1)}}{\sqrt{\cosh^2(\zeta)-e^{2i\lambda\omega t}\sinh^2(\zeta)}}
\end{split}
\end{equation}
where we used the analytic expression for the overlap of two squeezed states of the form: $\langle r | r e^{i\varphi} \rangle = ( \cosh^2(r) - e^{i\varphi} \sinh^2(r) )^{-1/2}$. 
The off-diagonal elements of the reduced clock state in Eq.\ \eqref{eq:FullStateClock} give all the information on the phase and visibility of the clock, through $2\rho_\mrm{eg}=V(t)e^{i\varphi(t)}$. We now give their explicit expressions to order $O(\varepsilon_\mrm{c}^2)$, as needed for the qSODS. The visibility of the clock, $ V = 2 | \rho_\mrm{eg} |$, is 
\begin{align}
\begin{split} 
    V
    &= \frac{\mrm{sech}^2(\zeta)}{\sqrt[4]{1 + \tanh^4(\zeta) - 2 \tanh^2(\zeta) \cos(2 \lambda \omega t)}} 
     \\ 
    &\simeq 1- \frac{\varepsilon_\mrm{c}^2}{16}\sin^2(\omega t) +O(\varepsilon_\mrm{c}^3) . \,  
\end{split}
\end{align}
The phase evolution of the clock is given by 
\begin{align}
    \varphi(t) &= \omega_\mrm{c} t +\frac{\omega t}{2}(\lambda-1) 
    \nonumber \\
    & \quad -\frac{1}{2}\arg\left(\cosh^2(\zeta)-e^{2i\lambda \omega t }\sinh^2(\zeta)\right)\\\nonumber
    & \simeq  \omega_\mrm{c} t -\frac{\omega t\varepsilon_\mrm{c}}{4}-\varepsilon_\mrm{c}^2\left(\frac{\omega t}{16}-\frac{\sin(2\omega t)}{32}\right)+O(\varepsilon_\mrm{c}^3) .
\end{align}

Rewritten as a fractional frequency shift 
in analogy to those stated in the main text, this becomes
\begin{equation}
    \frac{\Delta \nu}{\nu}= -\frac{\varepsilon_\mrm{m}}{4}-\frac{\varepsilon_\mrm{m}\varepsilon_\mrm{c}}{16}+ \frac{\hbar^2\nu\sin(2\omega t)}{32m^2c^4 t}.
\end{equation}

\clearpage

\title{Supplemental Material for ``Quantum signatures of proper time in optical ion clocks''}

\maketitle

\setcounter{page}{1}
\renewcommand{\theequation}{S\arabic{equation}}
\setcounter{equation}{0}

\onecolumngrid

\renewcommand{\theequation}{S\arabic{equation}}
\onecolumngrid
\subsection*{Protocol for Extracting the ``Quantum SODS'' Frequency Shift}

Here we outline a protocol to extract the contributions due to $\hat{S}(\hat{\varepsilon}_\mrm{c}/4)$ in the relativistic dynamics of the clock, given by Eq. (3) in the main text. These contributions result in an additional frequency shift which we call qSODS, but as we have discussed, the shift is too small 
to become visible when running the conventional clock protocol where the motional states are traced out, as it scales quadratically in $\varepsilon_\mrm{c}$. We therefore perform 
a measurement that projects  
the motional state onto a particular superposition of number states after the clock evolution has proceeded instead. A projection onto the superposition basis $( | 0 \rangle + | 2 \rangle )/\sqrt{2}$ 
extracts the interference and should in-principle give rise to a measurable frequency shift. However upon performing this measurement, the resulting phase offset is
\begin{align}
    \phi &= - \frac{\omega t \varepsilon_\mrm{c}}{4} + \frac{\varepsilon_\mrm{c}}{4\sqrt{2}} \sin \left( 2 \omega t (1 - \varepsilon_\mrm{c} /4 ) \right) 
\end{align}
The first term yields vSODS, while the second is a genuine quantum contribution arising due to $\hat U_\mrm{cm}$, however it rapidly oscillates at the trap frequency with zero mean (this arises from the factor $h(t)$ in Eq.\ 13).
 
The temporal resolution of state-of-the-art experiments is limited by the finite duration of the Ramsey pulses used to interfere the clock states, and thus this contribution will average to zero. To overcome this time-dependent averaging, we propose instead to drive the clock's motion with internal-state-dependent forces, which give rise to additional time-\textit{independent} shifts after interfering the motional states. Consider the scenario in which the clock evolves for time $t$ under Eq.\ (1), after which we apply the state-dependent displacement
\begin{align}
    \hat D(\beta) &= e^{-i\beta \hat Z_\mrm{c} \hat X}  
    \vt 
\end{align}
where $\hat Z_\mrm{c} = | e \rangle\langle e | - | g \rangle\langle g |  $ acts on the clock, $\beta = \chi T /\hslash$ for interaction energy $\chi$ and duration $T$, and $\smash{ \h X = ( \h a + \h a^\dagger ) /\sqrt{2} }$ is the position quadrature of the motion. The displacement results in an additional phase shift between the two parts of the state i.e.\ between the first term that gives rise to vSODS and the second term, which is the motional change due to $\hat U_\mrm{cm}$. Moreover, it induces additional transitions between different motional states. In particular, applying it to the state in Eq. (13) we obtain the contributions
\begin{align}
\begin{split} 
    \h D( \beta ) | 0 \rangle 
    &= e^{-\beta^2/4} \sum_{k=0}^{+\infty} \frac{(-i \beta \h Z_\mrm{c} )^k}{\sqrt{2^kk!}} | k \rangle 
    \\ 
    \h D (\beta ) | 2 \rangle 
    &= \frac{e^{-\beta^2/4}}{\sqrt{2}} 
    \left( \h a^\dagger - \frac{i\beta \h Z_\mrm{c}}{\sqrt{2}} \right)^{\! \! 2}
    \, \sum_{k=0}^{+\infty} \frac{(-i\beta \h Z_\mrm{c} )^k}{\sqrt{2^kk!}}|k\rangle  
\end{split} 
\end{align}
corresponding to a coherent state and displaced Fock state, respectively. Importantly, this gives us access to the full ladder of oscillator motional states and thereby the ability to project onto superpositions other than $( | 0 \rangle + | 2 \rangle ) /\sqrt{2}$. Keeping this final measurement state $| \vartheta \rangle$ generic for now, we obtain for the conditional state of the clock,

\begin{align}
    | \psi_\mrm{c}(t) \rangle &= 
    \frac{\vartheta_0^+}{\sqrt{2}} | g \rangle + \frac{e^{-i\omega_\mrm{c}(1 - \varepsilon_\mrm{m}/4)t}}{\sqrt{2}} \bigg( \vartheta_0^- - \frac{\vartheta_2^- \varepsilon_\mrm{c} h(t)}{4\sqrt{2}} \bigg) | e \rangle 
    \label{eq:displaceprojectmain}
\end{align}
where we have denoted $\smash{ \vartheta_k^\pm = \langle \vartheta | e^{\pm i \beta \h X} | k \rangle } $ and $| k \rangle$ is the $k$th Fock state. In particular the third term of Eq.\ (\ref{eq:displaceprojectmain}), which arises due to $\smash{\hat U_\mrm{cm}}$, is multiplied by the complex-valued $\vartheta_2^-$. 

In particular, choosing $| \vartheta \rangle = ( | 0 \rangle + | 1 \rangle ) /\sqrt{2}$, we obtain
\begin{align}
    \vartheta_0^\pm &= \frac{e^{- \beta^2/4}}{\sqrt{2}} \left( 1 \pm \frac{i\beta}{\sqrt{2}} \right) 
    \vt \\ 
    \vartheta_2^- &= -i \beta e^{-\beta^2/4} \bigg( 1 - \frac{i \beta}{2\sqrt{2}} - \frac{\beta^2}{4} \bigg) 
    \label{eqnu2} . 
\end{align}
The off-diagonal element of the clock to leading order in $\varepsilon_\mrm{c} \ll 1$ is given by,
\begin{align}
    2 \rho_\mrm{eg} &= e^{- i\omega_\mrm{c}t} e^{- i \mrm{arg} (2 + 2 \sqrt{2} i \beta - \beta^2 )} \left( 1 + i \left( \frac{\omega t \varepsilon_\mrm{c}}{4} + \frac{\beta \varepsilon_\mrm{c} \sin^2(\omega t)}{\sqrt{2}(2  +\beta^2 )} - \frac{\beta^2 \varepsilon_\mrm{c} \sin(2\omega t)}{8(2+\beta^2)} \left( 3 - \frac{\beta^2}{2} \right) \right) \right) 
    \vt 
    \label{eqs8}
\end{align}
Re-approximating $1 + i x \simeq e^{i x}$, we can subsequently read off the phase shift directly from Eq.\ (\ref{eqs8}), 
    \begin{align}
        \phi &= - \frac{\omega t \varepsilon_\mrm{c}}{4} + \mrm{arg} \big( 2 + 2 \sqrt{2} i \beta - \beta^2 \big) - \frac{\beta \varepsilon_\mrm{c}  \sin^2(\omega t)}{\sqrt{2}(2  + \beta^2)}
        + \frac{\beta^2\varepsilon_\mrm{c} \sin(2\omega t)}{8 (2 + \beta^2 )} \bigg( 3 - \frac{\beta^2}{2} \bigg) .
        \label{eqshiftf5}
    \end{align}
The first term in Eq.\ (\ref{eqshiftf5}) is the usual vSODS contribution, the second is a contribution that only depends on the displacement, while the fourth is a time-dependent term that averages to zero over the Ramsey time (recall this occurs due to the finite temporal resolution of any realistic experiment). Meanwhile the third term, while time-dependent, does not average to zero, and is what we refer to as the ``quantum SODS (qSDOS)'' phase shift, 
\begin{align}
    - \frac{\beta \varepsilon_\mrm{c} \sin^2(\omega t)}{\sqrt{2}(2 + \beta^2)} \xrightarrow[\mrm{time \: average}]{}  \phi_\mrm{qSODS} = - \frac{\beta \varepsilon_\mrm{c}}{2\sqrt{2}( 2 + \beta^2)} .
    \label{eq10qSODS}
\end{align}
This constant phase offset arises because the first and last terms of Eq.\ (\ref{eqnu2}) multiply the oscillatory factor $h(t) = 1 - e^{-2i\omega t(1-\varepsilon_\mrm{c}/4)}$. Without these imaginary terms, the resulting phase shift goes as $\sim \mrm{Im} ( h ( t ) ) = \mrm{sin} (2 \omega t ( 1 - \varepsilon_\mrm{c}/4 )$, whereas with these terms, one has $\sim \mrm{Im} ( i \mrm{Im} ( \vartheta_2^-) h ( t )  ) = 2 \mrm{Im} (\vartheta_2^- ) \mrm{sin}^2( \omega t ( 1 - \varepsilon_\mrm{c}/4)$. The dependence of Eq.\ (\ref{eq10qSODS}) on $\beta\varepsilon_\mrm{c}$ comes from interference between the first and third terms of Eq.\ (\ref{eq:displaceprojectmain}), and is 
thereby a genuine quantum signature of proper time evolution. We refer to this signature as ``quantum SODS,'' since in a similar way to entanglement between the clock and its motion, it can only arise through quantization of the proper time evolution, $\hat \tau = \tau ( \hat x, \hat p)$. We remark that the leading order probability of success for the measurement of the motional states is 
\begin{align} 
    P= \frac{e^{-\beta^2/2}}{2} \left( 1 + \frac{\beta^2}{2} \left( 1 - \frac{3\varepsilon_\mrm{c}}{8} - \frac{\beta^2 \varepsilon_\mrm{c}}{16} \right) \right) \simeq \frac{2 + \beta^2}{4} e^{-\beta^2/2} 
\end{align} 
Thus, the displacement cannot be much larger than $\beta \simeq 1$, with $P \sim 0.2$ at $\beta=2$.

\end{document}